\documentclass[a4paper]{jpconf}
\pdfoutput=1

\usepackage{etex} 				
\usepackage[utf8]{inputenc}
\usepackage{cmap}    			
\usepackage[T1]{fontenc}    	
\usepackage[english]{babel}
\usepackage{cite}				
\usepackage{url}
\usepackage[per-mode=symbol,bracket-unit-denominator=false,detect-all,load-configurations=binary]{siunitx}
\DeclareSIUnit\loc{LOC}
\DeclareSIUnit\permil{\textperthousand}
\newcommand{\ie}{i.\,e.\ }
\newcommand{\eg}{e.\,g.\ }

\usepackage{booktabs}			
\usepackage{graphicx}			
\usepackage{multirow}
\usepackage{tikz}
\usetikzlibrary{arrows,positioning,shapes,topaths,calc,fit,backgrounds,matrix,shadows,automata,patterns,decorations.pathmorphing,decorations.pathreplacing,decorations.text,circuits.logic.US,trees,mindmap}
\usepackage{microtype}			

\begin{document}
\title{Micro-CernVM: Slashing the Cost of Building and Deploying Virtual Machines}
\author{J~Blomer, D~Berzano, P~Buncic, I~Charalampidis, G~Ganis, G~Lestaris, R~Meusel, V~Nicolaou}
\address{CERN, CH-1211 Genève 23, Switzerland}
\ead{jblomer@cern.ch}

\begin{abstract}
	The traditional virtual machine building and and deployment process is centered around the virtual machine hard disk image. 
	The packages comprising the VM operating system are carefully selected, hard disk images are built for a variety of different hypervisors, and images have to be distributed and decompressed in order to instantiate a virtual machine. 
	Within the HEP community, the CernVM File System has been established in order to decouple the distribution from the experiment software from the building and distribution of the VM hard disk images.
	
	We show how to get rid of such pre-built hard disk images altogether. 
	Due to the high requirements on POSIX compliance imposed by HEP application software, CernVM-FS can also be used to host and boot a Linux operating system. 
	This allows the use of a tiny bootable CD image that comprises only a Linux kernel while the rest of the operating system is provided on demand by CernVM-FS. 
	This approach speeds up the initial instantiation time and reduces virtual machine image sizes by an order of magnitude. 
	Furthermore, security updates can be distributed instantaneously through CernVM-FS.
	By leveraging the fact that CernVM-FS is a versioning file system, a historic analysis environment can be easily re-spawned by selecting the corresponding CernVM-FS file system snapshot.
\end{abstract}

\section{Introduction}
\label{sec:intro}

Virtual machines provide a uniform and portable environment for high energy physics applications~\cite{hepvirt11}.
The same virtual machine can be used for both development and for deployment on cloud infrastructures so that very same operating system libraries that are used to develop physics algorithms are also used to execute them.
As such, virtual machines provide a way to escape the so called ``dependency hell'', \ie the problem that applications might not work or behave differently when using slightly different (versions of) system libraries.
Furthermore, virtual machines remove the need to port applications to different operating systems.
This is one of the key benefits when it comes to working with volunteers that provide free CPU cycles on a variety of platforms (Windows, Linux, Mac)~\cite{lhcathome12}.
Properly archived, virtual machines are also an important ingredient for the long-term preservation of experiment data because they contain the complex historic software stack necessary to interpret data files~\cite{dphep12}.

The hard disk resembling a virtual machine is distributed in the form of a hard disk image file.
There are two contradicting approaches to creating virtual machine images.
The first approach creates a hard disk image from a standard installation of an operating system.
Such images fit a large base of different applications and use cases, such as web servers, machines for interactive login, or graphical workstations.
The image size is rather large, typically a few gigabytes, and due to the many packages contained in an image, images quickly become outdated.

The second approach creates a \emph{virtual appliance}.
The virtual appliance comprises a ``just enough operating system'', that is a stripped down version of an operating system tailored to only one or few applications and use cases.
An image size of a virtual appliance can be as small as a few hundred megabytes. 

As we have seen maintaining the CernVM virtual appliance\footnote{\url{http://cernvm.cern.ch}}, even for a minimalistic image it can take hours to prepare package repositories, build and compress virtual machine images in various formats, and to test them.
This delay between making a change and verifying the result of the change restricts the speed at which the virtual appliance can be developed.
On the user's end, despite the fact that sophisticated tools have been developed to manage virtual machine images~\cite{vmfs07,qcow208,sheepdog10,imagedist11,swift12}, the image distribution problem is best mitigated by small images. 

With all its libraries, tools, and configuration data, an operating system is a complex software stack.
The problem of distributing complex and frequently changing software stacks to virtual machines were previously faced for experiment applications.
Experiment applications are now widely distributed by the CernVM File System (CernVM-FS)~\cite{cvmfs11}, a caching, read-only file system that downloads software bits and pieces on demand from an HTTP content delivery network.
Here, we bring the use of CernVM-FS to the next level and use it to distribute as well the operating system to a virtual machine.
The result is a tiny virtual machine image only consisting of a Linux kernel and the CernVM-FS client.
The actual operating system is then booted from CernVM-FS.
Instead of a ``just enough operating system'', this approach leads to an \emph{operating system on demand}.

The rest of the paper is structured as follows.
Section~\ref{sec:components} describes the components of a $\mu$CernVM system.
Section~\ref{sec:stack} describes the virtual machine's root file system stack and the connection between $\mu$CernVM and the operating system on CernVM-FS.
Section~\ref{sec:os} describes the maintenance of an operating system on CernVM-FS; in particular we focus on exploiting the internal versioning of CernVM-FS that allows the very same virtual machine image to instantiate any operating system ever released on CernVM-FS.
Section~\ref{sec:updates} describes how $\mu$CernVM based systems are updated.
Section~\ref{sec:related} compares $\mu$CernVM to similar approaches.

%
%
%
%

\section{Building blocks}
\label{sec:components}

\begin{figure}
	\caption{
		A $\mu$CernVM based virtual machine is twofold.
		The $\mu$CernVM image contains a Linux kernel, the AUFS union file system, and a CernVM-FS client.
		The CernVM-FS client connects to a special repository containing the actual operating system.
		The two CernVM-FS repositories contain the operating system and the experiment software.
	}
	\label{fig:components}
	\begin{center}
		\input{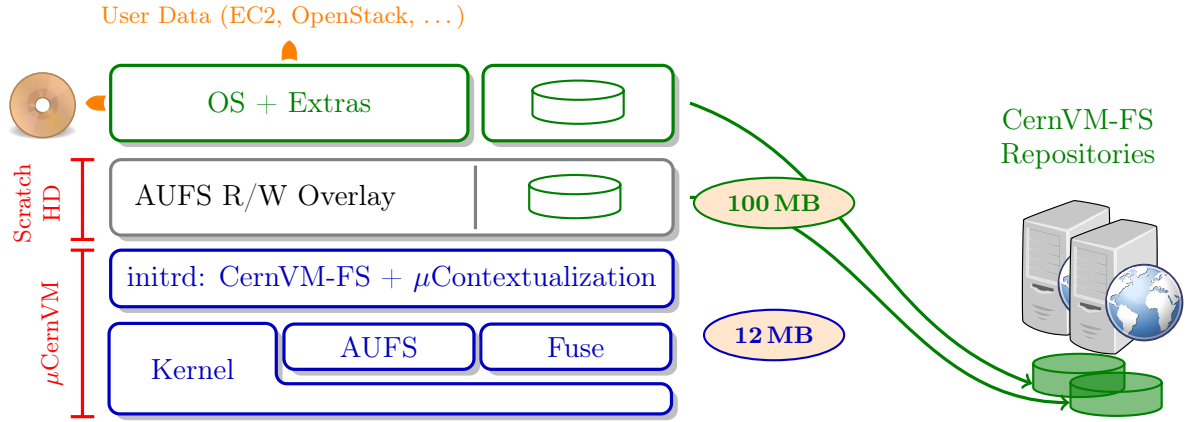}
	\end{center}
\end{figure}

Figure~\ref{fig:components} shows the key components of the $\mu$CernVM image.
The image consists of a Linux kernel and an init ramdisk that includes the CernVM-FS client.
The Linux kernel is ``virtualization friendly''.
It is slim as it only requires device drivers for the handful of hypervisors around.
Table~\ref{tab:kernel} compares the size of the $\mu$CernVM kernel with an Scientific Linux 6 kernel.
On the other hand, the $\mu$CernVM kernel has the latest para-virtualized drivers which are not necessarily found in Scientific Linux or other distribution kernels.

\begin{table} 
	\caption{\label{tab:kernel}Comparison of the size of the $\mu$CernVM kernel and the Scientific Linux 6 kernel}
	\begin{center}
		\begin{tabular}{lll}
	\br
 							& {\bf $\mu$CernVM} 		& {\bf Scientific Linux 6}	\\
	\mr
	Number of modules		& 100						& 2000						\\
	Size (incl. modules) 	& \SI{8}{\mega\byte}		& \SI{120}{\mega\byte}		\\
	\br
\end{tabular}
	\end{center}
\end{table}

Furthermore, the $\mu$CernVM kernel contains the \emph{union file system}~\cite{unionfs04} AUFS\footnote{\url{http://aufs.sourceforge.net}}.
The union file system is needed because CernVM-FS is a read-only file system.
Local changes, such as of files in /etc or /var, cannot be written to CernVM-FS.
Instead, the union file system transparently redirects such changes to a locally attached scratch hard disk.
The local hard disk is also used to store the CernVM-FS cache.
Note that the scratch hard disk does \emph{not} need to be distributed.
It can be created instantaneously when instantiating the virtual machine as an empty, sparse file.  

The init ramdisk contains the CernVM-FS client and a steering script.
The purpose of the steering script is to create the virtual machine's root file system stack that is constructed by unifying the CernVM-FS mount point with the writable scratch space.
To do so, the steering script can process contextualization information (sometimes called ``user data'') from various sources, such as OpenStack, OpenNebula, or Amazon~EC2.
Based on the contextualization information, the CernVM-FS repository and the repository version is selected.

The amount of data that needs to be loaded in order to boot the virtual machine is very little.
The image itself sums up to some \SI{12}{\mega\byte}.
In order to boot Scientific Linux 6 from CernVM-FS, the CernVM-FS client downloads additional \SI{100}{\mega\byte}.
The CernVM-FS infrastructure used to distribute experiment software can be reused.
In comparison, the (already small) CernVM~2.6 virtual appliance sums up to \SIrange{300}{400}{\mega\byte} that needs to be fully loaded and decompressed upfront before the boot process can start.
As a result, booting a $\mu$CernVM virtual machine starts practically instantaneously so that it can be, for instance, integrated with a web site that starts a virtual machine on the click of a button.\footnote{An example of such a web site is a volunteer computing project by the CERN theory group: \url{http://crowdcrafting.org/app/cernvm}}

\section{The $\mu$CernVM root file system stack}
\label{sec:stack}

At the beginning of the Linux boot process, in the so called \emph{early user space}, the Linux kernel uses a root file system in memory provided by the init ramdisk.
The purpose of the early user space is to load the necessary storage device drivers to access the actual root file system.
Once the actual root file system is available, the system switches its root file system to the new root mount point after which the previous root file system becomes useless and is removed from memory.

\begin{figure}
	\caption{\label{fig:rootfs}
		Transformation of the root file system stack.
		Left hand side: the root file system of the early user space.
		The /root directory is an AUFS mount point.
		Right hand side: the final root file system stack.
	}
	\begin{center}
		\scriptsize\begin{tikzpicture}
	[
		dirent/.style={
			circle,
			draw=green!50!black!50,
			fill=green!50!black!50
		},
		dirtree/.style={
			draw=gray
		},
		movement/.style={->,very thick,draw=red,
		},
		switch/.style={->,very thick,draw=blue,
		},
		process/.style={
			->,
			green!50!black!75,
			very thick
		},
		background/.style={
			rectangle,
			fill=gray!10,
			inner sep=0.2cm,
			rounded corners=5mm}
	]
	\node[dirent,label=right:/] (root) at (0,0) {};
	\node[dirent,label=right:root.rw (Scratch disk)] (rw) at (1, -1) {};
	\node[dirent,label=right:persistent] (persistent) at (2, -2) {};
	\node[dirent,label=right:cvmfs-cache] (cache) at (2, -3) {};
	\node[dirent,label=right:root.ro (CernVM-FS)] (ro) at (1, -4) {};
	\node[dirent,label=right:bin] (bin) at (2, -5) {};
	\node[dirent,label=right:etc] (etc) at (2, -6) {};
	\node at (2,-6.5) {\large$\vdots$};
	\node[dirent,label=right:root] (newroot) at (1, -7.5) {};
	
	\node[dirent,label=right:/] (realroot) at (9,0) {};
	\node[dirent,label=right:bin] (realbin) at (10,-1) {};
	\node[dirent,label=right:etc] (realetc) at (10,-2) {};
	\node[dirent,label=right:mnt] (realmnt) at (10,-3) {};
	\node[dirent,label=right:.rw] (realrw) at (11,-4) {};
	\node[dirent,label=right:.ro] (realro) at (11,-5) {};
	\node[dirent,label=right:cvmfs] (realcvmfs) at (10,-6) {};
	\node[dirent,label=right:atlas.cern.ch] (realatlas) at (11,-7) {};
	\node at (9.5,-7.5) {\large$\vdots$};
			
	\draw[dirtree] (root) -- (0,-1) -- (rw);
	\draw[dirtree] (rw) -- (1,-2) -- (persistent);
	\draw[dirtree] (1,-2) -- (1,-3) -- (cache);
	\draw[dirtree] (0,-1) -- (0,-4) -- (ro);
	\draw[dirtree] (ro) -- (1,-5) -- (bin);
	\draw[dirtree] (1,-5) -- (1,-6) -- (etc);
	\draw[dirtree] (0,-4) -- (0,-7.5) -- (newroot);
	
	
	\draw[dirtree] (realroot) -- (9,-1) -- (realbin);
	\draw[dirtree] (9,-1) -- (9,-2) -- (realetc);
	\draw[dirtree] (9,-2) -- (9,-3) -- (realmnt);
	\draw[dirtree] (realmnt) -- (10,-4) -- (realrw);
	\draw[dirtree] (10,-4) -- (10,-5) -- (realro);
	\draw[dirtree] (9,-3) -- (9,-6) -- (realcvmfs);
	\draw[dirtree] (realcvmfs) -- (10,-7) -- (realatlas);
	\draw[dirtree] (9,-6) -- (9,-7);
	
	\draw[movement,curve to,in=110,out=200] (persistent) to (newroot);
	\draw[movement,curve to,in=70,out=290] (ro) to (newroot);
	\node[red, fill=white] at ($(newroot) + (0,0.75)$) {AUFS};
	
	\draw[switch,curve to,out=-30,in=220] (newroot) to node[blue, fill=white, very near end] {switch root} (realroot);
	\draw[switch,curve to,out=-20,in=160] (rw) to node[blue, fill=white, near end] {bind mount} (realrw);
	\draw[switch,curve to,out=-20,in=190] (ro) to node[blue, fill=white, near end] {bind mount} (realro);
	

%
%
%
\end{tikzpicture}
	\end{center}
\end{figure}
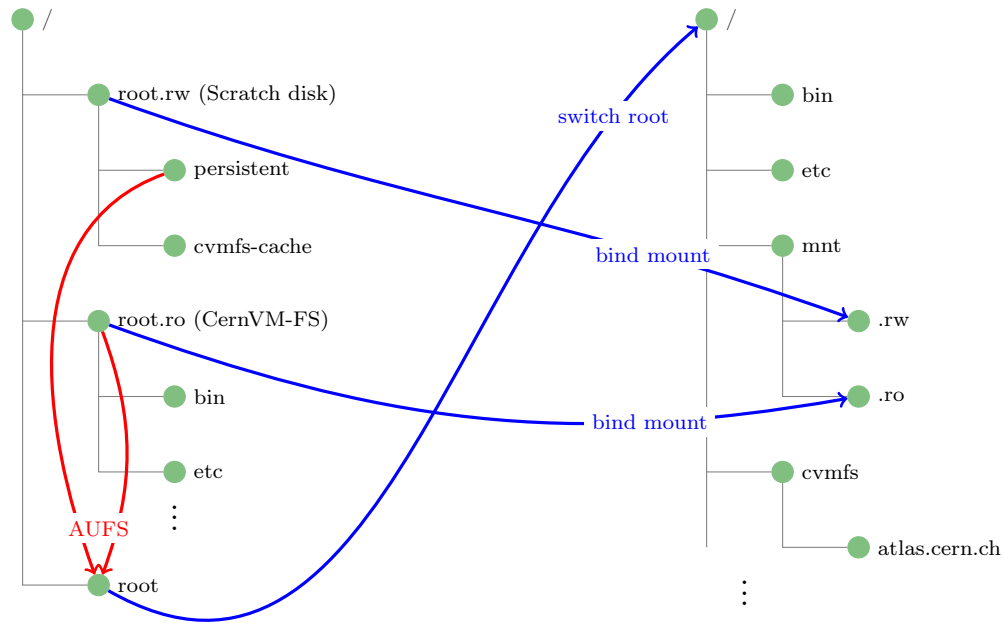

Figure~\ref{fig:rootfs} shows the transformation of the file system tree in the early user space in $\mu$CernVM.
First, the scratch hard disk is mounted on /root.rw.
$\mu$CernVM grabs the first empty hard disk or partition attached to the virtual machine, or remaining free space on the boot hard disk.
It automatically partitions, formats, and labels the scratch space.
Due to the file system label, $\mu$CernVM finds an already prepared scratch space on next boot.
The scratch space is used as a persistent writable overlay for local changes to the root file system and as a cache for the CernVM-FS client that loads the operating system.
Secondly, a CernVM-FS repository containing a template operating system installation is mounted on /root.ro.
The file system unification of the CernVM-FS mount point and the directory containing the persistent overlay is mounted on /root.
Finally, the /root directory is installed as new root file system.
The /root.ro and /root.rw directories are projected into the final root file system via \emph{bind mounts}.
The rest of the init ramdisk is removed from memory.

Although $\mu$CernVM and the operating system loaded by $\mu$CernVM are independent from each other, there are a few possible points of interaction between the two components.
These interaction points are implemented through special files in the root directory.
The following files are created or recognized by $\mu$CernVM:
\begin{description}
    \item[/.cvmfs\_pids] In this file, $\mu$CernVM stores the process IDs (PIDs) of the CernVM-FS process used to mount the operating system repository.
    	The PIDs help operating system scripts to avoid accidental killing of these vital processes.
	\item[/.ucernvm\_pinfiles]
		This file is provided by the operating system repository and contains a list of files that should be pinned in the cache.
		The idea is to always keep a minimal set of files available in the cache that allows for recovery from a broken network connection.
	\item[/.ucernvm\_bootstrap]
		This is a shell script provided by the operating system repository.
		It is sourced just before the root file system is switched and allows for custom actions.
\end{description}

Special care has to be taken in the shutdown script of the operating system.
Typically, the shutdown script does not expect an AUFS file system stack as root file system.
Therefore, the script needs to be modified to first remount the underlying read-write layer (/mnt/.rw) in read-only mode before remounting the root file system itself in read-only mode.
Only then the machine can be halted without risking any file system corruption.

\section{Versioned operating system on CernVM-FS}
\label{sec:os}

An easy way to provide an operating system template installation on CernVM-FS is to use the operating system's package manager to install the desired packages in the CernVM-FS repository area.\footnote{Like \texttt{yum --installroot /cvmfs/sl6.cern.ch install glibc emacs \dots}}
New and updated packages can then be installed incrementally by the package manager on top of the existing installation.
While being very fast, this results, however, in an ever-changing operating system directory tree.
It can quickly become difficult to trace back which update introduced or fixed a certain problem and which state of the operating system directory tree is mounted by virtual machines.

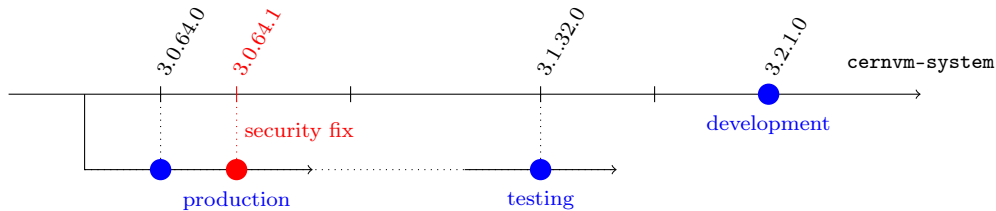
\begin{figure}
	\caption{\label{fig:version}
		CernVM 3 versioning scheme.
		Every released cernvm-system meta package is resembled as a versioned snapshot in CernVM-FS.
		The named branches \emph{development}, \emph{testing}, and \emph{production} provide entry points to versioned snapshots at different levels of maturity.
	}
	\begin{center}
		\scriptsize\begin{tikzpicture}
			\draw[->] (0,0) -- (12,0);		
			\draw[dotted] (1, -1) -- (8, -1);	
			\draw[->] (1,0) -- (1, -1) -- (4, -1);
			\draw[->] (6, -1) -- (8, -1);
			

			\draw (4.5cm,3pt) -- (4.5cm,-3pt);
			\draw (8.5cm,3pt) -- (8.5cm,-3pt);
			\draw[red] (3cm,3pt) -- (3cm,-3pt);
			
			\draw (2cm,3pt) -- (2cm,-3pt);
			\draw[dotted] (2,0) -- (2,-1);
			\draw (2,-1) node[circle, blue, fill=blue] {};
			\draw (3,-1) node[blue, anchor=north, yshift=-5pt] {\scriptsize production};
			\draw (2,0) node[anchor=west, yshift=5pt, xshift=-1pt, rotate=60] {\scriptsize 3.0.64.0};
			\draw (3,-1) node[circle, red, fill=red] {};
			\draw[dotted, red] (3,0) -- node[anchor=west, red] {\scriptsize security fix} (3,-1);
			\draw (3,0) node[anchor=west, yshift=5pt, xshift=-1pt, rotate=60, align=left, red] {\scriptsize 3.0.64.1};

			\draw (7cm,3pt) -- (7cm,-3pt);
			\draw[dotted] (7,0) -- (7,-1);
			\draw (7,-1) node[circle, blue, fill=blue] {};
			\draw (7,-1) node[blue, anchor=north, yshift=-5pt] {\scriptsize testing};
			\draw (7,0) node[anchor=west, yshift=5pt, xshift=-1pt, rotate=60] {\scriptsize 3.1.32.0};
			
			\draw (10,0) node[circle, blue, fill=blue] {};
			\draw (10,0) node[blue, anchor=north, yshift=-5pt] {\scriptsize development};
			\draw (10,0) node[anchor=west, yshift=5pt, xshift=-1pt, rotate=60] {\scriptsize 3.2.1.0};
			
			\draw(12,0) node[anchor=south, yshift=5pt] {\scriptsize\tt cernvm-system};

		\end{tikzpicture}

	\end{center}
\end{figure}

We have tackled this problem for \emph{CernVM 3}, a Scientific Linux 6 based operating system on CernVM-FS.
CernVM 3 benefits from versioning on three levels that ensure traceability and a well-defined, predictable virtual machine (see Figure~\ref{fig:version}).
\begin{enumerate}
	\item The \emph{cernvm-system} meta package provides the notion of a well defined operating system.
		The meta package has no payload but contains only package dependencies.
		In case of the cernvm-system package, the dependencies are fully versioned and no dependencies are missing, \ie there is no degree of freedom for the dependency resolution of this package.
	\item Three different CernVM-FS repositories are maintained, \emph{development}, \emph{testing}, and \emph{production}, that reflect different levels of maturity of the cernvm-system meta package.
		Furthermore, separation of repositories allows for injection of a security hotfix in the production repository while continue to develop in the other repositories.
	\item To avoid silent and unwanted updates of the virtual machine operating system, $\mu$CernVM exploits the fact that CernVM-FS is a versioning file system.
		CernVM-FS creates a snapshot whenever changes are published.
		These snapshots remain available and they can be named.
		The concept and the implementation is similar to a \emph{tag} in the git versioning system.
		CernVM-FS clients can, through mount options, select a particular snapshot on mount.\footnote{This feature is so far untapped by experiments' usage of CernVM-FS.  Instead experiments use different directories for different software releases, reflecting the fact that many versions of the experiment software need to be available on the same worker node at the same time.}
		
		On first boot, $\mu$CernVM mounts the newest available snapshot of the given repository and it will stay on this snapshot during reboots unless instructed otherwise.
		Snapshots are named after the version number of the cernvm-system meta package.
		Contextualization can be used to select another snapshot to mount, letting $\mu$CernVM go back in time and instantiate a historic data processing environment.
\end{enumerate}

The three levels of versioning allow the $\mu$CernVM image to always instantiate the newest available stable operating system version.
At the same time, the very same image provides instant access to every other operating system version ever released in the given repository.

\section{System updates}
\label{sec:updates}

While some virtual machines are short-lived and only instantiated for a particular computing job, others can run for many weeks or months.
A typical example of a long-lived virtual machine is an interactive virtual machine on an end user's laptop.
It is desirable to provide a long-lived virtual machine with bug fixes and security updates without the need to reconfigure the virtual machine or to replace the image file.
This section describes how to update both kernel and init ramdisk on $\mu$CernVM and the operating system on CernVM-FS.

\subsection{Updates of $\mu$CernVM}
The $\mu$CernVM image can be used as a read-only CD-ROM image.
Hence writing updated data to the image is not possible.
However, updated versions of the Linux kernel and the init ramdisk can be dropped into a predefined location on the scratch hard disk.
Upon boot, after mounting the scratch hard disk, $\mu$CernVM uses the \emph{kexec} facility~\cite{kexec03} in order to reboot on the fly into the updated kernel and init ramdisk.

\subsection{Updates of an operating system on CernVM-FS}
Once booted, a $\mu$CernVM based virtual machine can be kept up to date by standard means of the package manager provided by the operating system.
This approach, however, accumulates more and more local changes on the writable overlay stored on the scratch hard disk.
Over time, more and more disk space is spent on the writable overlay and the local operating system diverges from the one provided on CernVM-FS.

Instead it is preferable to pick up updates from CernVM-FS.
To this end, $\mu$CernVM can be instructed to ``unpin'' the currently mounted CernVM-FS snapshot and to mount the newest available snapshot on next boot.
Changing the snapshot can result in conflicts with the local changes on the writable overlay.
For instance, a user might have replaced /usr/bin/gcc while at the same time the updated snapshot provides a newer version of /usr/bin/gcc as well.
The conflict resolution in $\mu$CernVM is guided by the widely used file system hierarchy standard~\cite{fhs04}: $\mu$CernVM resolves conflicts in /etc and /var by keeping the locally modified versions, whereas in all other directories the version from CernVM-FS has precedence.
Furthermore, packages installed by the user need to be reinserted into the package database of the updated snapshot.
Finally, the system account databases (/etc/passwd, /etc/group, \dots) are merged record by record as follows
\begin{itemize}
	\item Local passwords have precedence
	\item Group membership is merged
	\item Conflicting user ids or group ids from the new snapshot are mapped to the locally defined ids.
		If necessary, user ids or group ids from the new snapshot are remapped to previously unused ids.
		Remapping of user ids and group ids is supported by CernVM-FS, in the same way NFS clients support id mappings.~\cite{rfc1813}
\end{itemize}
This approach to updates assumes that operating system components comply with common standards (for instance, that components do not store configuration data in /usr/bin).
While first observations with CernVM 3 are positive, it requires more experience to verify the approach in practice.

\section{Related Work}
\label{sec:related}

With live CDs and network booted Linux systems, $\mu$CernVM shares the fact that the root file system is read-only and local changes are written to an overlay area by means of a union file system.
Live CDs need to be loaded as a whole before the boot process can start and the writable overlay is in memory.
Typical network boot setups rely on NFS, which cannot be used for wide-area setups.
Also, booting from the network relies on a network boot protocol (\eg PXE) instead of using a minimal virtual machine image.

An idea very similar to $\mu$CernVM was implemented as \emph{HTTP-FUSE Xenoppix}~\cite{httpfuse06}.
This system consists of a kernel and an init ramdisk that downloads the actual root file system on demand via HTTP.
Important features of CernVM-FS are not implemented though, such as versioning and digitally signed network traffic.
Unlike $\mu$CernVM, HTTP-FUSE Xenoppix runs only on the Xen hypervisor.
There is no discussion of system updates in HTTP-FUSE Xenoppix.

The \emph{vagrant} tool set greatly facilitates building images for virtual appliances, especially ``virtual development environments''~\cite{vagrant12}.
Despite the convenient user interface, hard disk images containing an full operating system must still be distributed.

Similarly, \emph{CoreOS}\footnote{\url{http://coreos.com}} simplifies massive server deployment.
It consists of a tiny Linux system just enough to run lightweight virtual machines (``containers'') but it is not a virtual machine by itself.
Instead of loading files on demand, CoreOS manages entire containers.

\section{Conclusion}
We have presented $\mu$CernVM, a \SI{12}{\mega\byte} virtual machine image that loads the actual operating system files on demand through CernVM-FS.
Updates to the virtual machine do not require changing the image but they are simply distributed through changes in the CernVM-FS repository.
Moreover, the very same tiny image can boot any version of the operating system ever released on CernVM-FS, which facilitates the long-term preservation of data processing environments.

We have verified our approach by creating CernVM~3, the SL6 based successor of CernVM~2, on top of $\mu$CernVM.\footnote{The CernVM~3 build system is available under \url{https://github.com/cernvm}.}
CernVM~3 runs on VMware, VirtualBox, Xen, KVM, and Hyper-V hypervisors as well as in OpenStack, OpenNebula, and Amazon EC2 clouds.
CernVM~3 can build itself and it runs experiment software from the four LHC experiments.
A systematic validation of experiment software remains to be done.

\section*{References}
\bibliography{bibliography}

\providecommand{\newblock}{}
\begin{thebibliography}{10}
\expandafter\ifx\csname url\endcsname\relax
  \def\url#1{{\tt #1}}\fi
\expandafter\ifx\csname urlprefix\endcsname\relax\def\urlprefix{URL }\fi
\providecommand{\eprint}[2][]{\url{#2}}

\bibitem{hepvirt11}
Buncic P, Sanchez C~A, Blomer J, Harutyunyan A and Mudrinic M 2011 {\em The
  European Physical Journal Plus\/} {\bf 126}

\bibitem{lhcathome12}
Gonz{\'a}lez D~L, Grey F, Blomer J, Buncic P, Harutyunyan A, Marquina M, Segal
  B and Skands P 2012 {\em PoS(ISGC 2012)036\/}

\bibitem{dphep12}
{The ICHFA DPHEP International Study Group} 2012 Status report of the dphep
  study group: Towards a global effort for sustainable data preservation in
  high energy physics Tech. Rep. FERMILAB-PUB-12-878-PPD Fermilab

\bibitem{vmfs07}
{VMware} 2007 {VMware} virtual machine file system: Technical overview and best
  practices Tech. Rep. WP-022-PRD-01-01 VMware

\bibitem{qcow208}
McLoughlin M 2008 The {QCOW2} image format
  \url{https://people.gnome.org/~markmc/qcow-image-format.html}

\bibitem{sheepdog10}
Morita K 2010 {Sheepdog}: Distributed storage system for {QEMU}/{KVM} talk at
  the LinuxCon Japan 2010

\bibitem{imagedist11}
Wartel R, Cass T, Moreira B, Roche E, Guijarro M, Goasguen S and Schwickerath U
  2010 {\em Proc. of the 2nd IEEE Int. Conf. on Cloud Computing Technology and
  Science (CloudCom'10)\/} pp 112--117

\bibitem{swift12}
{SwiftStack} 2012 The openstack object storage system Tech. rep. {SwiftStack}

\bibitem{cvmfs11}
Blomer J, Aguado-Sanchez C, Buncic P and Harutyunyan A 2011 {\em Journal of
  Physics: Conference Series\/} {\bf 331}

\bibitem{unionfs04}
Wright C~P, Dave J, Gupta P, Krishnan H, Zadok E and Zubair M~N 2004
  Versatility and unix semantics in a fan-out unification file system Tech.
  Rep. FSL-04-01b Stony Brook University

\bibitem{kexec03}
Pfiffer A 2003 Reducing system reboot time with kexec Tech. rep. Open Source
  Development Labs

\bibitem{fhs04}
Russell R, Quinlan D and Yeoh C 2004 Filesystem hierarchy standard Tech. rep.
  freestandards.org

\bibitem{rfc1813}
Callaghan B, Pawlowski B and Staubach P 1995 {NFS Version 3 Protocol
  Specification} RFC 1813 Internet Engineering Task Force
  \urlprefix\url{http://www.rfc-editor.org/rfc/rfc1813.txt}

\bibitem{httpfuse06}
Suzaki K, Yagi T, Iijima K, Kitagawa K and Tashiro S 2006 {\em Proc. of the
  2006 Linux Symposium\/} vol~2 pp 379--392

\bibitem{vagrant12}
Palat J 2012 {\em Linux Journal\/}

\end{thebibliography}

\end{document}